\documentclass[journal,twoside,web]{ieeecolor}
\usepackage{generic}
\usepackage{cite}
\usepackage{url}
\usepackage{amsmath,amssymb,amsfonts}
\usepackage{algorithm}
\usepackage{algorithmic}
\usepackage{graphicx}
\usepackage{textcomp}
\usepackage{tabularx}
\usepackage{hyperref}
\usepackage{orcidlink}
\usepackage[skip=1pt,compatibility=false]{caption}
\usepackage{xcolor}

\usepackage{multirow}
\usepackage{booktabs}
\usepackage{makecell}
\usepackage[normalem]{ulem} 

\def\BibTeX{{\rm B\kern-.05em{\sc i\kern-.025em b}\kern-.08em
    T\kern-.1667em\lower.7ex\hbox{E}\kern-.125emX}}
\begin{document}
\bstctlcite{IEEEexample:BSTcontrol}
\title{Diff-DDoS: Realistic Cyber-Physical Attack Synthesis and Robust Detection for 5G-Enabled CPS Using Tabular Diffusion Models}

\author{
Bilal Hussain\orcidlink{0000-0002-0046-9007}, \IEEEmembership{Senior Member, IEEE},
Xiao Tang\orcidlink{0000-0001-8971-5413}, \IEEEmembership{Member, IEEE},
Qinghe Du\orcidlink{0000-0002-6656-6061}, \IEEEmembership{Member, IEEE},
Tan Li\orcidlink{0000-0001-6129-4792}, \IEEEmembership{Member, IEEE},
Muhammad Azhar\orcidlink{0000-0003-3687-0270}, \IEEEmembership{Member, IEEE},
and Danista Khan\orcidlink{0000-0002-8450-5991}, \IEEEmembership{Member, IEEE}
\thanks{Received March 18, 2026; revised July 6, 2026; accepted August 11, 2026. This work was fully supported by a grant from the College of Professional and Continuing Education, an affiliate of The Hong Kong Polytechnic University. Paper no. TII-26-6533. \textit{(Corresponding author: Bilal Hussain.)}}
\thanks{Bilal Hussain is with the Division of Science, Engineering, and Health Studies, School of Professional Education and Executive Development, The Hong Kong Polytechnic University, Hong Kong SAR, China (e-mail: bilal.hussain@cpce-polyu.edu.hk).}
\thanks{Xiao Tang and Qinghe Du are with the School of Information and Communications Engineering, Xi'an Jiaotong University, Xi'an, China (e-mails: tangxiao@xjtu.edu.cn; duqinghe@mail.xjtu.edu.cn).}
\thanks{Tan Li is with the Department of Computer Science, The Hang Seng University of Hong Kong, Hong Kong SAR, China (e-mail: tanli@hsu.edu.hk).}
\thanks{Muhammad Azhar is with the Department of Applied Data Science, Hong Kong Shue Yan University, Hong Kong SAR, China (e-mail: azhar@hksyu.edu).}
\thanks{Danista Khan is with the Division of Business and Hospitality Management, School of Professional Education and Executive Development, The Hong Kong Polytechnic University, Hong Kong SAR, China (e-mail: danista.khan@cpce-polyu.edu.hk).}
\thanks{Digital Object Identifier XX.XXXX/TII.2026.XXXXXXX}
\thanks{This is the authors' accepted manuscript. Accepted for publication in \textsc{IEEE Transactions on Industrial Informatics}, paper no.\ TII-26-6533. \copyright~2026 IEEE. Personal use of this material is permitted. Permission from IEEE must be obtained for all other uses, in any current or future media, including reprinting/republishing this material for advertising or promotional purposes, creating new collective works, for resale or redistribution to servers or lists, or reuse of any copyrighted component of this work in other works.}
}

\maketitle

\begin{abstract}
Deep learning-based DDoS detectors for 5G-enabled cyber-physical systems face two challenges: scarce real-world labeled attack data and the unrealism of naive synthetic substitutes, which limit robustness against adaptive adversaries. 
Detectors trained on hand-crafted attacks with fixed scaling multipliers degrade catastrophically---F1-score drops of ${\sim}47\%$ to $100\%$ depending on scenario---when confronted with realistic, distribution-preserving samples. 
We propose \emph{Diff-DDoS}, a three-phase framework for realistic attack synthesis and robust detection using tabular diffusion models. Phase~1 establishes a baseline CNN cell-level detector on spatiotemporal grids derived from call detail records~(CDRs). 
Phase~2 trains a tabular denoising diffusion probabilistic model~(TabDDPM) on normal CDR aggregates to generate realistic attacks, exposing detector vulnerabilities. 
Phase~3 introduces adversarial diffusion training~(ADT), which applies inverse classifier guidance to iteratively generate hard yet distribution-preserving adversarial samples until the detector converges. 
On a real-world Milano CDR dataset across SMS-flooding, silent-call, Internet-signaling, and blended scenarios, ResNet50 with ADT recovers F1-scores of $79.62\%$ (silent-call), $100\%$ (Internet), and $92.79\%$ (blended) while retaining near-perfect multiplier-test F1 on most scenarios.
We further benchmark ADT against gradient-based adversarial training, conditional tabular GAN (CTGAN), and fixed-multiplier baselines under identical iterative training and validation-based threshold calibration; after calibration, ADT reaches $100\%$ SMS F1 versus $47.3\%$ for CTGAN and attains silent-call F1 on par with the strongest gradient-based adversarial-training baseline.
SMS flooding remains challenging for lightweight architectures, though calibration enables near-perfect ResNet50 detection.
These results establish tabular diffusion models as a practical tool for stress-testing and hardening intrusion detectors in data-scarce 5G cyber-physical deployments.
\end{abstract}


\section{Introduction}
\label{sec:introduction}
\IEEEPARstart{F}{ifth}-generation (5G) mobile networks are becoming a core communication substrate for cyber-physical systems (CPSs), including industrial monitoring, transportation, and energy services.
This convergence enlarges the attack surface and creates new pathways for distributed denial-of-service (DDoS) attacks that threaten both network availability and physical processes~\cite{Humayed2017, He2018, 3gpp2017tr33899, Hussain2021}, requiring multi-layer AI-enabled protection~\cite{alam2026, hussain2026closedloop}.
Recent threat-landscape reports rank DDoS attacks among the most prominent cybersecurity threats in mobile and Internet of Things (IoT)-connected environments~\cite{enisa2025}.

Detecting \emph{stealthy} cellular attacks is particularly difficult.
Silent-call abuse, SMS flooding, and Internet-signaling (signaling storm) attacks stress control-plane procedures rather than only consuming user-plane bandwidth~\cite{Tu2015, Murynets2013, nguyen2025rrc, wen2024oran}, and signaling overload and layer-3 abuse remain credible threats in modern cellular architectures~\cite{nguyen2025rrc, wen2024oran}.

At the monitoring layer, call detail record (CDR) aggregates capture cell-level temporal deviations in SMS, call, and Internet activity. Hussain \emph{et al.}~\cite{Hussain2021} showed they can be organized as $9\times9\times3$ spatiotemporal grids and classified by convolutional neural networks (CNNs) with strong performance on multiplier-based synthetic attacks.
However, those multiplier attacks use fixed, large feature multipliers and are therefore easier to separate from normal traffic than realistic adversarial behavior.

This mismatch exposes a broader weakness: models trained on stylized attacks degrade sharply under distribution shift~\cite{he2023adversarial, alatwi2021adversarial}. In cellular settings the problem is amplified by scarce, privacy-sensitive attack labels; naive perturbations can break traffic semantics~\cite{He2018,Hussain2021,he2023adversarial,alatwi2021adversarial}, leaving detectors brittle against attacks that stay near the normal traffic manifold.

To address this gap, we propose Diff-DDoS, a three-phase framework using tabular diffusion models to synthesize realistic attack samples and harden detectors against adaptive evasion.
In \textbf{Phase~1}, we re-implement the CNN-based detector of~\cite{Hussain2021} as a multiplier-based baseline (full pseudo-code in Supplementary Sec.~S-A).
In \textbf{Phase~2}, we train TabDDPM only on normal CDR aggregates to generate manifold-preserving attack samples, revealing the baseline detector's poor generalization.
In \textbf{Phase~3}, we introduce \emph{adversarial diffusion training} (ADT), which uses guided diffusion sampling to generate hard on-manifold attacks and iteratively retrains the detector.

The main contributions of this work are as follows:
\begin{enumerate}
\item We re-implement the cell-level CNN baseline of~\cite{Hussain2021} and establish reference performance on multiplier-based attacks.
\item We demonstrate that detectors trained only on multiplier-based attacks can fail severely when evaluated on diffusion-generated attacks drawn from the normal CDR manifold.
\item We introduce ADT, which combines TabDDPM-based generation with inverse classifier guidance to create hard, realistic attack samples for iterative robust training.
\item We evaluate Diff-DDoS across SMS, silent-call, Internet, and blended scenarios using SimpleCNN and 50-layer residual network (ResNet50), highlighting ADT robustness gains, validation-based threshold calibration, and a per-scenario learnability analysis showing that SMS flooding is capacity-limited on SimpleCNN but recoverable on ResNet50.
\end{enumerate}

Sec.~II covers related work; Sec.~III background and problem setup; Sec.~IV methodology; Sec.~V experimental setup and evaluation protocols; Sec.~VI results; Sec.~VII limitations and conclusion. Extended derivations and full result grids are in the Supplementary Material.

\vspace{-4pt}
\section{Related Work}\label{sec:related_work}
DDoS detection in 5G-enabled CPSs has advanced from rule-based to machine learning/deep learning (ML/DL) methods because signaling abuse, SMS flooding, and silent-call attacks can evade threshold defenses~\cite{Mavoungou2016, He2018}. Hussain \emph{et al.}~\cite{Hussain2021} showed CDR grids enable strong CNN performance on multiplier-based attacks; later work explored recurrent, federated, and quantum-inspired models~\cite{dahiya2023ddos, munaweera2024federated, swathi2025securing}.

CDR analytics also demonstrate useful spatiotemporal structure for traffic prediction and network monitoring~\cite{naboulsi2016large, xu2017understanding, furno2017city, naboulsi2014classifying}. Yet both DDoS detection and CDR analytics literatures rarely evaluate robustness to adaptive evasion close to normal traffic statistics~\cite{sultan2018call, Aziz2024CDR, Hussain2021, dahiya2023ddos}.

Deep intrusion detection systems (IDSs) degrade under distribution shift and adversarial manipulation~\cite{shone2018deep, apruzzese2018effectiveness, arp2022dos}. 
Conditional tabular generative adversarial network (CTGAN) improves data augmentation but suffers from instability~\cite{xu2019modeling}; TabDDPM offers richer tabular distributions and higher-fidelity synthesis~\cite{kotelnikov2023tabddpm, Stoian2026}. 
Gradient-based adversarial perturbations generate off-manifold samples~\cite{goodfellow2015explaining, madry2018towards, he2023adversarial, alatwi2021adversarial}, while diffusion-based approaches show growing promise~\cite{Alauthman2026GANIDS, nie2022diffusion, merzouk2025diffusionids}.

\section{Background and Problem Setup}
\label{sec:background}
\subsection{CDR Grid Representation and Detection Task} \label{sec:problem_formulation}
Each CDR snapshot is represented as a grid image $x \in \mathbb{R}^{9 \times 9 \times 3}$, where the three channels correspond to aggregated SMS-out, Call-out, and Internet traffic for each of the $9 \times 9 = 81$ cells in a selected Milano~\cite{barlacchi2015multi} subgrid over one 10-minute interval within the $11{:}00$--$14{:}00$ window.
The detection task is formulated as a multilabel cell-level classification problem: for each grid image $x$, a CNN-based detector $\mathcal{M}$ with learnable weights $\theta$ outputs a vector
\[
\mathcal{M}(x;\theta) = \hat{Y} \in (0,1)^{81},
\]
where each component $\hat{y}_k \in (0,1)$ is the sigmoid-activated predicted attack probability for cell $k \in \{1,\dots,81\}$. 
The ground-truth label $Y \in \{0,1\}^{81}$ assigns $Y_k = 1$ if cell $k$ is under attack and $Y_k = 0$ otherwise. 
During adversarial diffusion training (Phase~3), $\theta$ is updated iteratively across rounds, so the detector after ADT round $r$ for attack scenario $s$ is denoted $\mathcal{M}(\cdot\,;\theta_s^{r})$ (during round-$r$ sampling, guidance uses $\theta_s^{r-1}$; Sec.~\ref{sec:phase3}).
We write $f_\theta \equiv \mathcal{M}(\cdot;\theta)$ in algorithms when shorthand is needed.

The CDR-based attack synthesis of~\cite{Hussain2021} applies scenario-specific multipliers to $|\mathcal{C}|=40$ randomly selected cells, labeling them $Y_k=1$ (see Supplementary Algorithm~S2). 
Four scenarios are evaluated throughout: \textit{SMS flooding} (elevated SMS-out), \textit{silent-call} (suppressed outgoing calls), \textit{Internet-signaling} (elevated Internet, approximating a signaling storm~\cite{Hussain2021}), and \textit{blended} (simultaneous SMS/Internet elevation and call suppression). 
Phase~1 uses fixed multipliers ($169.5\times$ SMS, $-33.1\times$ call clipped at zero, $43.3\times$ Internet); these stylized attacks motivate the diffusion-based robustness evaluation and are \emph{not} used to generate attacks in Phases~2--3 (Phase~2 test attacks are TabDDPM-generated; see Sec.~\ref{tabddpm_demo}).
The preprocessing pipeline (62 CDR files, $11{:}00$--$14{:}00$ aggregation, $\mathrm{Num2Dims}(\cdot)$ row-major cell indexing) is given in Supplementary Algorithms~S1--S2.

\subsection{Denoising Diffusion and Classifier Guidance} \label{subsec:ddpm_prelim}
We use standard denoising diffusion probabilistic model (DDPM) notation~\cite{ho2020denoising,kotelnikov2023tabddpm}: a fixed forward schedule corrupts clean data $x_0$ into noisy samples $x_t$ over $T$ steps; a learned reverse denoiser $\epsilon_\psi$ (weights $\psi$, distinct from detector $\theta$) predicts the injected noise. 
Our TabDDPM instantiation on 3D CDR aggregates is summarized in Sec.~\ref{subsec:tabddpm_training}; full equations are in Supplementary Sec.~S-B.
DDPMs have been extended from images~\cite{dhariwal2021diffusion} to tabular data~\cite{kotelnikov2023tabddpm}; Diff-DDoS builds on the latter.

Classifier guidance~\cite{dhariwal2021diffusion} steers the reverse process via classifier gradients:
\begin{equation}
\nabla_{x_t} \log p(x_t \mid y)
= \underbrace{\nabla_{x_t} \log p(x_t)}_{\text{unconditional score}}
+ s_{\mathrm{cg}} \cdot \underbrace{\nabla_{x_t} \log p(y \mid x_t)}_{\text{classifier gradient}},
\end{equation}
where $\nabla_{x_t}$ is the gradient with respect to the noisy sample $x_t$, $s_{\mathrm{cg}}$ controls conditioning strength, and the classifier gradient biases samples toward target class $y$.
Diff-DDoS extends this idea via \emph{inverse} guidance in Phase~3 (Sec.~\ref{sec:phase3}), steering samples away from attack-like detector scores while preserving the normal CDR manifold~\cite{nie2022diffusion,merzouk2025diffusionids}.

\subsection{Notation Summary} \label{subsec:notation}
Table~\ref{tab:notation} lists overloaded symbols used in the main text; the supplement provides an extended disambiguation table.

\vspace{2pt}
\begin{table}[t]
\centering
\caption{Main-text notation summary.}
\label{tab:notation}
\footnotesize
\setlength{\tabcolsep}{2.5pt}
\renewcommand{\arraystretch}{0.98}
\begin{tabularx}{\linewidth}{@{}l>{\raggedright\arraybackslash}X@{}}
\toprule
\textbf{Symbol} & \textbf{Meaning} \\
\midrule
$\mathcal{M}(\cdot;\theta)$, $f_\theta$ & Multilabel CNN detector ($81$ sigmoid outputs per grid) \\
\cmidrule(lr){1-2}
$\Xi_s^{r}$, $\Xi_s^{r,\mathrm{orig}}$, $\Xi_s^{r,\mathrm{Tab}}$ & Metric bundles (Acc, Prec, Rec, F1, FPR) for attack scenario $s$ at ADT round $r$; \emph{not} the detector \\
\cmidrule(lr){1-2}
$\theta$, $\theta_s^{*}$, $\theta_s^{r}$ & Detector learnable weights \\
\cmidrule(lr){1-2}
$\epsilon_\psi$, $\phi_{\mathrm{TabDDPM}}$ & TabDDPM denoiser network (weights $\psi$; distinct from $\theta$) \\
\cmidrule(lr){1-2}
$\phi_s:\mathbb{R}^3\!\to\!\mathbb{R}^3$ & Mild post-sample scenario scaling on 3D CDR aggregates (Phase~2) \\
\cmidrule(lr){1-2}
$\mathrm{MapToGrid}(\tilde{\mathbf{x}},B)$ & Copy background $B$, write aggregate $\tilde{\mathbf{x}}$ into attacked cells $\mathcal{C}$, set $Y_k=1$ on $\mathcal{C}$ \\
\cmidrule(lr){1-2}
$p_{\mathrm{attack}}$ & Mean detector score over attacked cells $\mathcal{C}$ during inverse guidance \\
\bottomrule
\end{tabularx}
\vspace{-3pt}
\end{table}

\begin{figure*}[t]
\centering
\includegraphics[width=0.95\textwidth]{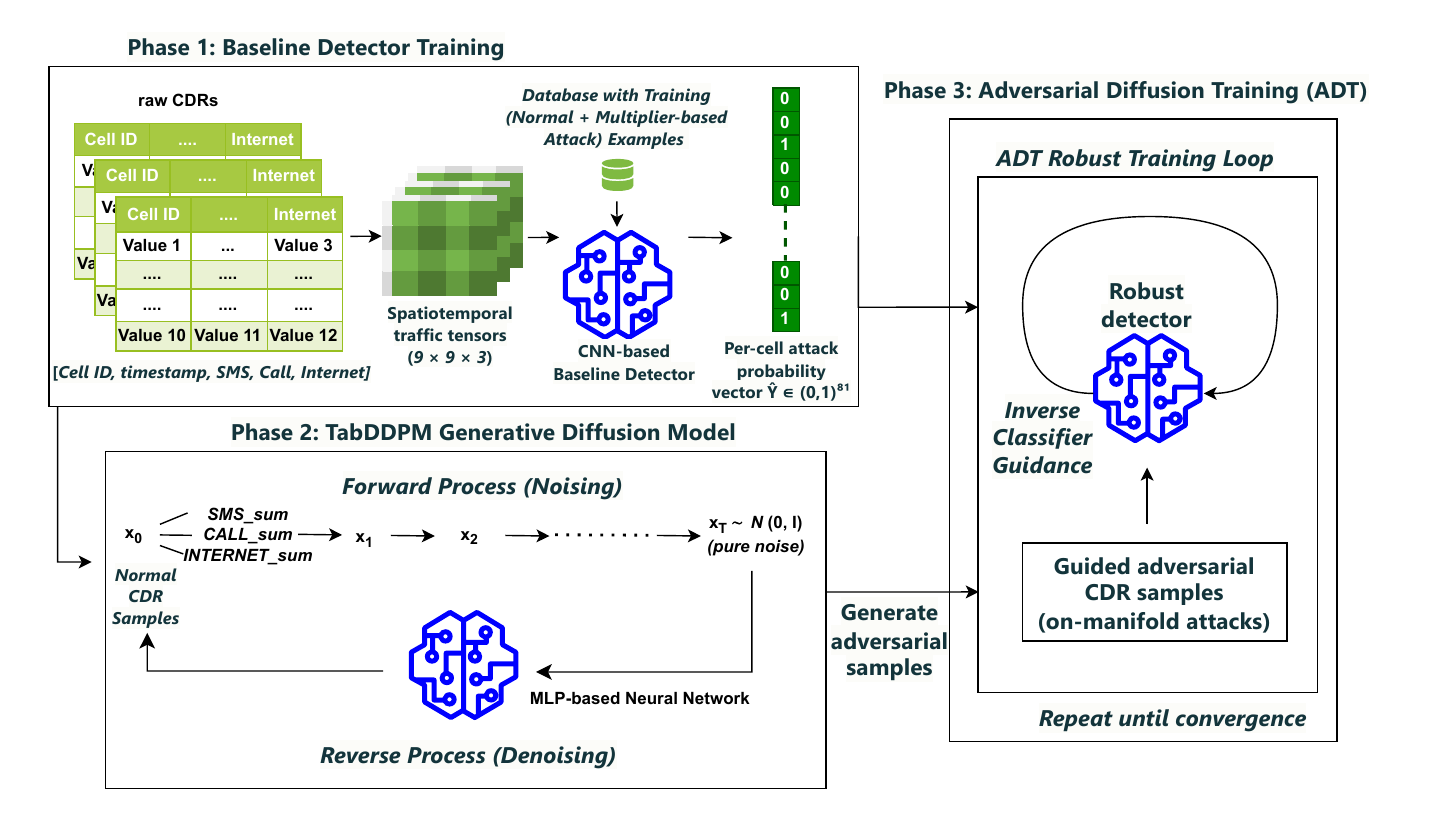}
\vspace{-4pt}
\caption{Overview of the proposed Diff-DDoS framework. Phase~1 trains baseline detectors on multiplier-based attacks, Phase~2 uses TabDDPM to generate realistic attacks from normal CDR aggregates, and Phase~3 applies ADT to generate hard on-manifold samples for iterative retraining.}
\label{fig:system_model}
\vspace{-12pt}
\end{figure*}

\section{Methodology}
\label{sec:methodology}
Fig.~\ref{fig:system_model} summarizes the Diff-DDoS pipeline: \textbf{Phase~1} establishes a multiplier-trained CNN baseline; \textbf{Phase~2} trains TabDDPM on normal 3D CDR aggregates and exposes baseline vulnerability to manifold-preserving attacks; \textbf{Phase~3} applies adversarial diffusion training (ADT) with inverse classifier guidance to iteratively harden the detector.
Offline attack synthesis is separated from online grid-level inference (Sec.~\ref{subsec:inference_latency}).

\subsection{Data Representations and Attack-Steering Mechanisms} \label{subsec:data_representations}
Normal tabular training rows are $\mathcal{R}_{\mathrm{norm}}=\{(\mathrm{SMS\_sum},\mathrm{CALL\_sum},\mathrm{INTERNET\_sum})\}$ extracted from raw CDR aggregates over the $9\times9$ subgrid; they are standardized with \texttt{StandardScaler} fit on $\mathcal{R}_{\mathrm{norm}}$ only.
$\mathrm{Num2Dims}(k)$ maps cell index $k$ to grid coordinates $(h,w)$ in row-major order.
$\mathrm{MapToGrid}(\tilde{\mathbf{x}},B)$ copies background grid $B$, writes aggregate $\tilde{\mathbf{x}}$ into attacked cells $\mathcal{C}$ via $\mathrm{Num2Dims}(\cdot)$, and sets $Y_k=1$ on $\mathcal{C}$.
Phase~2 \emph{test} grids use $B\in\mathcal{A}$ (abnormal-template pool); 
Phase~3 ADT \emph{training} grids use a zero background ($B=\mathbf{0}$): only cells in $\mathcal{C}$ receive $\tilde{\mathbf{x}}$; all others remain zero (Algorithm~\ref{alg:phase3}).
Table~\ref{tab:steering} contrasts the three attack-steering mechanisms used across phases.

\begin{table}[t]
\centering
\caption{Attack-steering mechanisms across Diff-DDoS phases.}
\label{tab:steering}
\footnotesize
\setlength{\tabcolsep}{3pt}
\begin{tabular}{@{}p{0.22\linewidth}p{0.14\linewidth}p{0.28\linewidth}p{0.26\linewidth}@{}}
\toprule
\textbf{Mechanism} & \textbf{Phase} & \textbf{Action} & \textbf{Purpose} \\
\midrule
Fixed multipliers & 1 & $169.5\times$ SMS, $-33.1\times$ Call, $43.3\times$ Internet & Stylized baseline \\
Heuristic steering $+$ $\phi_s$ & 2 & Mild $\gamma$ ranges on target channel(s) & On-manifold vulnerability test \\
Inverse classifier guidance & 3 & Minimize $p_{\mathrm{attack}}$ toward $p_{\mathrm{target}}$ & Hard robust-training samples \\
\bottomrule
\end{tabular}
\end{table}

\vspace{-4pt}
\subsection{Phase~1: Baseline Detector Training} \label{subsec:phase1_training}
The detector is a \emph{multilabel} model with 81 sigmoid outputs per grid (Sec.~\ref{sec:problem_formulation}); Phase~1 trains on the union of normal grids (all-negative labels) and multiplier-attack grids (40 perturbed cells positive), without any TabDDPM-generated attacks.
Phase~1 follows~\cite{Hussain2021}; the full training procedure is in Supplementary Algorithm~S3.

\textbf{Data Preprocessing:} Phase~1 uses the two-stage pipeline of Supplementary Algorithms~S1--S2 (grid extraction and fixed-multiplier attack injection on $|\mathcal{C}|=40$ cells; Table~\ref{tab:steering}). Attack labels are defined solely by synthesis; no statistical thresholding is applied.

\textbf{Model Training:} Two detector architectures are employed:
\begin{itemize}
    \item \textbf{SimpleCNN:} A lightweight CNN with three 
    convolution--pooling blocks followed by fully-connected layers.
    \item \textbf{ResNet50:} A deeper (50-layer) residual network 
    adapted to $9 \times 9 \times 3$ inputs with an 81-way 
    sigmoid output layer.
\end{itemize}

Both models are trained with a multilabel binary cross-entropy (BCE) loss averaged over $N$ training images and all 81 cells:
\begin{equation}
\begin{aligned}
\mathcal{L}_{\mathrm{BCE}}(\theta)
&= -\frac{1}{N} \sum_{i=1}^{N} \frac{1}{81} \sum_{k=1}^{81}
\Bigl[
y_{i,k} \log \hat{y}_{i,k} \\
&\qquad\qquad\qquad
+ (1-y_{i,k}) \log(1-\hat{y}_{i,k})
\Bigr],
\end{aligned}
\end{equation}
where $y_{i,k} \in \{0,1\}$ is the ground-truth label and $\hat{y}_{i,k} = \mathcal{M}(x_i;\theta)_k \in (0,1)$ is the sigmoid-activated predicted attack probability for cell $k$ in image $i$, produced by detector $\mathcal{M}$ with learnable weights $\theta$.

Predictions are binarized at $\tau=0.5$; true positives, false positives, and false negatives (TP/FP/FN) are micro-averaged over $N_{\mathrm{test}}\times81$ cell decisions to compute Accuracy, Precision, Recall, F1 score, False Positive Rate (FPR), and receiver operating characteristic area under the curve (ROC--AUC) (Sec.~\ref{subsec:roc_auc}; dataset splits in Sec.~\ref{sec:experimental_setup}).

\vspace{-4pt}
\subsection{Phase 2: TabDDPM-Based Vulnerability Demonstration} \label{tabddpm_demo}
Phase~2 demonstrates that multiplier-trained detectors are vulnerable to realistic diffusion-based attacks.
Let $\mathcal{X}_n$ denote the $781$ normal-only grid images ($614$ train $+$ $167$ test) and $\mathcal{A}$ the $335$ unperturbed abnormal-template grids produced by Supplementary Algorithms~S1--S2 (fixed random seed~$1$; Sec.~\ref{sec:experimental_setup}).
Let $\mathcal{X}_a^{s}$ denote the scenario-specific abnormal grid working pool ($N_{\mathrm{abn}}=335$) used in Phase~2 indexing.
Training attacks in $\mathcal{X}_a^{s}$ (indices $0$--$166$) remain multiplier-based, while test attacks (indices $167$--$334$, $N_{\mathrm{attack}}^{\mathrm{test}}=168$) are replaced with TabDDPM-generated grids.
Let $\mathcal{X}_a^{s,\mathrm{Tab}}$ denote the TabDDPM-based test set: the $335$-image split combining $167$ held-out normal test grids with $168$ TabDDPM test-attack grids $\mathcal{X}_a^{s}[167{:}335]$ (i.e., $\mathcal{D}_{\mathrm{test}}^{(s)}$ in Algorithm~\ref{alg:phase2-tabddpm-vulnerability}).
Phase-2 F1 is $\mathrm{F1}_{\mathrm{Ph.2}}^{(s)}$ on this split with $\tau=0.5$ (protocol~P1; Table~\ref{tab:protocols}); degradation is $\Delta_{\mathrm{F1}}^{(s)}=(\mathrm{F1}_{\mathrm{Ph.1}}^{(s)}-\mathrm{F1}_{\mathrm{Ph.2}}^{(s)})/\mathrm{F1}_{\mathrm{Ph.1}}^{(s)}\times 100\%$.

\subsubsection{Tabular Diffusion Model for Aggregated CDRs}
\label{subsec:tabddpm_training}
Attack synthesis operates on 3D CDR aggregates (SMS, CALL, INTERNET) per attacked-cell group, where scenario constraints are naturally expressed and realism is quantified (Sec.~\ref{subsec:realism_metrics}).
A grid-image DDPM would couple unrelated cells and complicate per-channel control; we therefore model the low-dimensional aggregate manifold with TabDDPM~\cite{kotelnikov2023tabddpm} and map sampled aggregates back to $9\times9\times3$ grids via $\mathrm{MapToGrid}(\cdot,B)$ (Sec.~\ref{subsec:data_representations}).

We adapt the DDPM framework~\cite{ho2020denoising} to tabular CDR features. Raw CDR rows are aggregated into a 3D feature vector $x_0 = (\mathrm{SMS\_sum},\mathrm{CALL\_sum},\mathrm{INTERNET\_sum})\in\mathbb{R}^3$ per $(\text{cell},\text{timeslot},\text{day})$, which defines the normal CDR manifold. A fixed linear noise schedule ($T=1000$, $\beta_{\min}=10^{-4}$, $\beta_{\max}=0.02$) governs the forward diffusion, and the denoiser $\epsilon_\psi$---a multilayer perceptron (MLP) with hidden dimensions $[256,512,256]$ that takes the concatenation of $x_t$ and the scalar diffusion step $t$ (broadcast to the feature dimension)---is trained with the standard $\ell_2$ noise-prediction objective $\mathcal{L}_{\mathrm{DDPM}}(\psi)$.
The full forward/reverse equations, denoiser architecture, training objective, and posterior sampling are given in Supplementary Sec.~S-B.

\subsubsection{Realistic Attack Scenario Generation}
TabDDPM is trained exclusively on normal aggregate CDR rows. 
Here $\mathcal{U}(a,b)$ denotes the continuous uniform distribution on $[a,b]$, and $\gamma_1,\gamma_2,\gamma_3$ are independent per-sample scaling factors.
To generate realistic attacks for a given scenario $s$, we:
\begin{enumerate}
    \item Run reverse diffusion (Supplementary Sec.~S-B) to obtain an aggregate $\mathbf{x}^{(i)}=(\mathrm{SMS},\mathrm{CALL},\mathrm{INTERNET})\in\mathbb{R}^3$, applying lightweight heuristic steering on the target channel(s) every $\max(1,\lfloor T/10\rfloor)$ timesteps (Table~\ref{tab:steering}; distinct from Phase~3 inverse classifier guidance and from post-sample $\phi_s$ in item~2).

    \item Apply mild scenario-specific constraints $\tilde{\mathbf{x}}^{(i)} \leftarrow \phi_s(\mathbf{x}^{(i)})$ on the target channel only: SMS, $\gamma_1 \sim \mathcal{U}(2,5)$; silent-call, $\gamma_2 \sim \mathcal{U}(0.3,0.7)$ with $\mathrm{CALL}\leftarrow\max(0,\gamma_2\cdot\mathrm{CALL})$; Internet, $\gamma_3 \sim \mathcal{U}(2,5)$; Blended, all three applied jointly.
    Non-target channels are left unchanged (silent-call models resource-exhaustion via call suppression~\cite{Tu2015, Xie2018}).

    \item Map each constrained triple to a $9 \times 9 \times 3$ grid via $\mathrm{MapToGrid}(\tilde{\mathbf{x}}^{(i)},B^{(i)})$ with $B^{(i)}\in\mathcal{A}$.
\end{enumerate}
The Phase~1 detectors are then evaluated on the full test split without retraining (results in Sec.~\ref{subsec:phase_progression}).

\begin{algorithm}[t]
\caption{Phase 2: TabDDPM Vulnerability Demonstration}
\label{alg:phase2-tabddpm-vulnerability}
\begin{algorithmic}[1]
\REQUIRE Supplementary Algorithms~S1--S2 outputs
  ($\mathcal{D}_{\mathrm{norm}}$, $\mathcal{A}$, attacked-cell set $\mathcal{C}$);
  detector $\mathcal{M}(\cdot;\theta)$ (written $f_\theta$)
\ENSURE Baseline $\{f_{\theta_s^*}\}_{s}$; $\{\mathrm{F1}_{\mathrm{Ph.2}}^{(s)}, \Delta_{\mathrm{F1}}^{(s)}\}_{s}$

\STATE Build $\mathcal{R}_{\mathrm{norm}}$ from normal CDR aggregates; fit \texttt{StandardScaler} on $\mathcal{R}_{\mathrm{norm}}$
\STATE Train TabDDPM $p_\psi$ on scaled $\mathcal{R}_{\mathrm{norm}}$ (Supplementary Sec.~S-B); save model and scaler

\FOR{each $s \in \{\mathrm{SMS},\mathrm{Call},\mathrm{Internet},\mathrm{Blended}\}$}
  \STATE Train $f_\theta$ on scenario-specific $\mathcal{D}_s^{\mathrm{train}}$
    (Algorithm~S2; no TabDDPM in training) $\Rightarrow f_{\theta_s^*}$
  \STATE Initialize $\mathcal{X}_a^{s}$ with multiplier attacks on all $N_{\mathrm{abn}}=335$ slots
  (indices $0$--$166$ retained for training)
  \FOR{$i=167$ to $334$ \COMMENT{$N_{\mathrm{attack}}^{\mathrm{test}}=168$ test-attack indices}}
    \STATE $\mathbf{x}^{(i)} \leftarrow \text{TabDDPM-Sample}(p_\psi)$ with heuristic scenario steering on target channel(s) during reverse diffusion every $\max(1,\lfloor T/10\rfloor)$ steps (enumerate item~1; not Phase~3 inverse classifier guidance)
    \STATE $\tilde{\mathbf{x}}^{(i)} \leftarrow \phi_s(\mathbf{x}^{(i)})$
      \COMMENT{post-sample mild $\gamma$ ranges in enumerate item~2}    
    \STATE $B^{(i)} \leftarrow \mathcal{A}[i]$;
      $\mathcal{X}_a^{s}[i] \leftarrow \mathrm{MapToGrid}(\tilde{\mathbf{x}}^{(i)}, B^{(i)})$
  \ENDFOR
  \STATE Assemble $\mathcal{D}_{\mathrm{test}}^{(s)}$ from $167$ held-out normal grids
    and $168$ TabDDPM attack grids $\mathcal{X}_a^{s}[167{:}335]$
    ($335$ test images total)
  \STATE Evaluate $f_{\theta_s^*}$ on $\mathcal{D}_{\mathrm{test}}^{(s)}$ without retraining;
    record $\mathrm{F1}_{\mathrm{Ph.2}}^{(s)}$, $\Delta_{\mathrm{F1}}^{(s)}$
\ENDFOR
\RETURN $\{f_{\theta_s^*}\}_{s}$, $\{\mathrm{F1}_{\mathrm{Ph.2}}^{(s)}, \Delta_{\mathrm{F1}}^{(s)}\}_{s}$
\end{algorithmic}
\end{algorithm}

\vspace{-4pt}
\subsection{Phase~3: Adversarial Diffusion Training (ADT)} \label{sec:phase3}
Phase~3 applies ADT: for each $s\in\mathcal{S}=\{\mathrm{SMS},\mathrm{Call},\mathrm{Internet},\mathrm{Blended}\}$ and round $r=1,\dots,R$, inverse-guided TabDDPM samples are appended to the training set and the detector is fine-tuned (Algorithm~\ref{alg:phase3}).
Let $\mathcal{X}_{\mathrm{adv}}^{r}$ denote the adversarial grids generated in round $r$ and $\phi_{\mathrm{TabDDPM}}$ the Phase~2 TabDDPM denoiser (Supplementary Sec.~S-B).

\subsubsection{Inverse Classifier Guidance}
During each ADT round, inverse classifier guidance steers TabDDPM sampling \emph{away} from attack-like detector scores.
Let $\mathcal{M}(\cdot;\theta_s^{r})$ denote the detector \emph{after} ADT round $r$ ($\theta_s^{0}=\theta_s^{*}$ from Phase~1). During generation at round $r$, inverse guidance uses $\mathcal{M}(\cdot;\theta_s^{r-1})$ (Algorithm~\ref{alg:phase3}); we omit subscript $s$ in Eqs.~(\ref{eq:p_attack})--(\ref{eq:guided_update}) when $s$ is fixed.
For a CDR aggregate $x$ (or its clean estimate $\hat{x}_0$ during reverse diffusion), we form the grid $I_{\mathrm{grid}}(x)=\mathrm{MapToGrid}(x,B)$ by replicating $x$ into all cells of $\mathcal{C}=\{c_1,\dots,c_{40}\}$ on a background grid $B$ (for inverse guidance during ADT sampling, $B$ is drawn from normal training grids; see ADT iterative training below).
The scalar attack probability is the mean predicted score over $\mathcal{C}$:
\begin{equation} \label{eq:p_attack}
p_{\mathrm{attack}}
= \frac{1}{|\mathcal{C}|}
\sum_{k \in \mathcal{C}}
\mathcal{M}(I_{\mathrm{grid}}(x);\theta_s^{r-1})_k.
\end{equation}
Given target probability $p_{\mathrm{target}}=0.3$ and guidance scale $\lambda$, the guidance term is
\begin{equation} \label{eq:guidance}
\nabla_{\mathrm{guidance}} := \lambda \cdot (p_{\mathrm{attack}} - p_{\mathrm{target}}) \cdot \nabla_x\, p_{\mathrm{attack}},
\end{equation}
where $\nabla_x\, p_{\mathrm{attack}}$ is the gradient of $p_{\mathrm{attack}}$ with respect to $x$; the guided update is
\begin{equation} \label{eq:guided_update}
\hat{x}_0 \leftarrow \hat{x}_0 - \nabla_{\mathrm{guidance}}.
\end{equation}
For non-blended scenarios, gradient components in non-target CDR channels are scaled by $0.7$ before applying the update, preventing guidance from distorting unrelated traffic dimensions.

\subsubsection{ADT Iterative Training Procedure}

\begin{algorithm}[!t]
\caption{Phase 3: adversarial diffusion training (ADT)}
\label{alg:phase3}
\begin{algorithmic}[1]
\REQUIRE Phase~1 weights $\{\theta_s^{*}\}_{s\in\mathcal{S}}$; TabDDPM $\phi_{\mathrm{TabDDPM}}$ (Supplementary Sec.~S-B);
  $(\mathcal{X}_s^{\mathrm{train}},\mathcal{Y}_s^{\mathrm{train}})$, $\mathcal{X}_s^{\mathrm{test}}$, $\mathcal{X}_a^{s,\mathrm{Tab}}$ (Phases~1--2);
  rounds $R$, guidance $\lambda$, samples/round $N_{\mathrm{ADT}}$, fine-tune epochs $E_{\mathrm{ADT}}$, cell set $\mathcal{C}$
\ENSURE $\{\theta_s^{R,*}\}_{s}$; $\{\Xi_s^{r}\}_{r=0}^{R}$

\FOR{each $s \in \mathcal{S}=\{\mathrm{SMS},\mathrm{Call},\mathrm{Internet},\mathrm{Blended}\}$}
  \STATE $\theta_s^{0}\gets\theta_s^{*}$;
    $\Xi_s^{0}\gets\mathrm{Evaluate}(\theta_s^{0},\mathcal{X}_a^{s,\mathrm{Tab}})$
  \FOR{$r=1$ to $R$}
    \STATE $\mathcal{X}_{\mathrm{adv}}^{r}\gets\emptyset$;
      $f_{\mathrm{guid}}\gets\max(1,\lfloor T/10\rfloor)$
    \FOR{$i=1$ to $N_{\mathrm{ADT}}$}
      \STATE TabDDPM reverse diffusion from $x_T\sim\mathcal{N}(0,I)$; at steps $t\bmod f_{\mathrm{guid}}=0$, $t>0$, apply inverse guidance (Eqs.~\eqref{eq:p_attack}--\eqref{eq:guided_update}) using $\mathcal{M}(\cdot;\theta_s^{r-1})$
      \STATE $x_{\mathrm{adv}}\gets\mathrm{MapToGrid}(\hat{x}_0,B{=}0)$
        \COMMENT{zero background; aggregate in $\mathcal{C}$ only}
      \STATE $\mathcal{X}_{\mathrm{adv}}^{r}\gets\mathcal{X}_{\mathrm{adv}}^{r}\cup\{x_{\mathrm{adv}}\}$    
    \ENDFOR
    \STATE Append $\mathcal{X}_{\mathrm{adv}}^{r}$ to $\mathcal{X}_s^{\mathrm{train}}$ (attack labels) $\Rightarrow(\mathcal{X}_s^{\mathrm{train},r},\mathcal{Y}_s^{\mathrm{train},r})$
    \STATE Fine-tune $\theta_s^{r-1}$ for $E_{\mathrm{ADT}}$ epochs $\Rightarrow\theta_s^{r}$
    \STATE $\Xi_s^{r}\gets\{\Xi_s^{r,\mathrm{orig}},\Xi_s^{r,\mathrm{Tab}}\}$ on $\mathcal{X}_s^{\mathrm{test}}$, $\mathcal{X}_a^{s,\mathrm{Tab}}$
  \ENDFOR
  \STATE $\theta_s^{R,*}\gets\theta_s^{R}$
\ENDFOR
\RETURN $\{\theta_s^{R,*}\}_{s}$, $\{\Xi_s^{r}\}_{r=0}^{R}$
\end{algorithmic}
\end{algorithm}

Round-wise metrics are $\Xi_s^{r}=\{\Xi_s^{r,\mathrm{orig}},\Xi_s^{r,\mathrm{Tab}}\}$ on the multiplier test split $\mathcal{X}_s^{\mathrm{test}}$ and TabDDPM test set $\mathcal{X}_a^{s,\mathrm{Tab}}$, respectively; the final weights are $\theta_s^{R,*}$.
During inverse guidance, $I_{\mathrm{grid}}(x)$ is formed with a randomly drawn \emph{normal} training background.
ADT-appended training grids use a zero background with the aggregate written only into $\mathcal{C}$ (Sec.~\ref{subsec:data_representations}).
Hyperparameters ($R$, $N_{\mathrm{ADT}}$, $E_{\mathrm{ADT}}$, $\lambda$, $p_{\mathrm{target}}$) are given in Sec.~\ref{sec:experimental_setup}.

\section{Experimental Setup} \label{sec:experimental_setup}
Following~\cite{Hussain2021}, we use the Telecom Italia Milano CDR collection~\cite{barlacchi2015multi}: a $9\times9$ subgrid ($81$ cells), 18 time slots per day ($11{:}00$--$14{:}00$, 10-minute intervals), yielding $1{,}116$ grid images per scenario split into $781$ training ($614$ normal $+$ $167$ attack) and $335$ test ($167$ normal $+$ $168$ attack) images. 
We fix random seed~$1$ for data shuffling and attack-cell selection ($|\mathcal{C}| = 40$).

TabDDPM is trained on normal aggregate CDR rows (3D features) for $100$ epochs (learning rate $10^{-4}$, batch size $32$). 
SimpleCNN and ResNet50 are trained for up to $100$ epochs in Phases~1--2 (ResNet50: $10\%$ validation split with early stopping to limit overfitting; SimpleCNN: full $100$-epoch budget).
Phase~3 ADT uses $R=3$ rounds, $\lambda=1.0$, $N_{\mathrm{ADT}}=100$ samples per round and scenario, $E_{\mathrm{ADT}}=20$ fine-tuning epochs per round, and $p_{\mathrm{target}}=0.3$.
We evaluate Accuracy, Precision, Recall, F1, FPR, and ROC~AUC via micro-averaging over $N_{\mathrm{test}}\times81$ cell decisions ($N_{\mathrm{test}}=335$).
\textbf{No normalization} is applied to CNN inputs (consistent with~\cite{Hussain2021}); TabDDPM aggregates are standardized with \texttt{StandardScaler} fit on normal data only.
Implementation details follow Supplementary Algorithms~S1--S3 and main-text Algorithms~1--2.

\subsection{Evaluation Protocols for Reported F1}
\label{subsec:eval_protocols}
F1 is reported under protocols P1--P5; Table~\ref{tab:protocols} maps each label to its evaluation setting.

\vspace{2pt}
\begin{table}[t]
\centering
\caption{Reader's guide to F1 evaluation protocols.}
\label{tab:protocols}
\footnotesize
\setlength{\tabcolsep}{3pt}
\begin{tabular}{@{}p{0.06\linewidth}p{0.38\linewidth}p{0.48\linewidth}@{}}
\toprule
\textbf{ID} & \textbf{Protocol} & \textbf{Main location} \\
\midrule
P1 & $335$-grid TabDDPM test; $\tau=0.5$; cumulative Phase-3 ADT & Table~\ref{tab:phases} \\
P2 & Same test as P1; fresh 3-round schedule per method; $\tau=0.5$ & Fig.~\ref{fig:baseline_threshold} (left) \\
P3 & Locked attack test; $\tau_{\mathrm{F1max}}$ or $\tau_{\mathrm{Y}}$ from validation & Fig.~\ref{fig:baseline_threshold} (right); Sec.~\ref{subsec:calibration} \\
P4 & Fresh-retrain $\lambda$/$p_{\mathrm{target}}$ sweep; locked test & Table~\ref{tab:guidance_sensitivity} \\
P5 & Multiplier test split; $\tau=0.5$ & Table~\ref{tab:multiplier_retention} \\
\bottomrule
\end{tabular}
\end{table}

\section{Experimental Results and Discussion} \label{sec:results}
Extended derivations, sensitivity sweeps, and full result grids are in the Supplementary Material.

\subsection{Evaluation Scope} \label{subsec:eval_scope}
Our use of ``5G'' refers to the \emph{deployment context}---CPS services depending on cellular connectivity---rather than a claim the benchmark traces were collected from a 5G standalone (SA) core.
We evaluate on the public Milano CDR corpus~\cite{barlacchi2015multi} for reproducible comparison with Hussain \emph{et al.}~\cite{Hussain2021}; coarse CDR aggregates remain generation-agnostic monitoring records used in contemporary 5G anomaly detection~\cite{aziz2024tmc,Aziz2024CDR}.
No public labeled per-cell cellular DDoS CDR dataset exists; packet-/flow-level 5G corpora (5G-NIDD~\cite{siriwardhana2025data}, NCSRD-DS-5GDDoS~\cite{ncsrd2024_5gddos}) are not directly mappable to our $9\times9\times3$ grid representation.
Empirical claims are scoped to robust CDR-grid detection under realistic synthetic stress tests in a 5G-motivated CPS monitoring setting.

\subsection{Phase Progression Across Diff-DDoS} \label{subsec:phase_progression}
Table~\ref{tab:phases} summarizes detector performance across phases (all values at $\tau=0.5$; Ph.~2--3 follow protocol~P1 on the TabDDPM test split).
Both architectures achieve near-perfect Phase~1 F1 ($99.6\%$--$100\%$) on multiplier-based test attacks---expected because Phase~1 multipliers (Table~\ref{tab:steering}) deviate tens to hundreds of $\sigma$ from normal (Fig.~\ref{fig:realism_zscore}).
Without retraining, Phase~2 F1 collapses to near zero for SMS, Internet, and Blended on TabDDPM attacks; ResNet50 silent-call F1 falls to $7.6\%$ despite greater model capacity (a comparatively deeper backbone).
ADT (Phase~3) recovers ResNet50 F1 to $79.6\%$ (silent-call), $100\%$ (Internet), and $92.8\%$ (Blended) at round~3; SMS F1 stays below $10\%$ at $\tau=0.5$ (a decision-rule artifact resolved in Sec.~\ref{subsec:calibration}).
SimpleCNN silent-call F1 declines under ADT (Sec.~\ref{subsec:gradcam}); for these sparse stealth scenarios (SMS flooding and silent-call), ResNet50 is the deployable backbone (Secs.~\ref{subsec:calibration}--\ref{subsec:gradcam}).

\begin{table}[t]
\centering
\caption{Detector performance across phases (protocol~P1; all values in \%; deployment-default $\tau=0.5$).
\textbf{Ph.~1:} multiplier-based test attacks (Acc, Rec, F1); precision is $100\%$ for all rows (omitted).
\textbf{Ph.~2:} TabDDPM-generated attacks without retraining.
\textbf{Ph.~3:} ADT rounds R1--R3.
SC: SimpleCNN; R50: ResNet50.}
\label{tab:phases}
\small
\setlength{\tabcolsep}{4pt}
\resizebox{1\columnwidth}{!}{%
\begin{tabular}{l|ccc|c|ccc}
\hline
\multirow{2}{*}{\textbf{Scenario}} &
\multicolumn{3}{c|}{\textbf{Ph.~1}} &
\multicolumn{1}{c|}{\textbf{Ph.~2}} &
\multicolumn{3}{c}{\textbf{Ph.~3 F1}} \\
\cline{2-8}
 & \textbf{Acc} & \textbf{Rec} & \textbf{F1} & \textbf{F1} &
 \textbf{R1} & \textbf{R2} & \textbf{R3} \\
\hline
SMS (SC)      & 99.96 & 99.85 & 99.93 & 0     & 9.09  & 9.09  & 9.09  \\
SMS (R50)\textsuperscript{\textbf{a}}     & 100   & 100   & 100   & 0     & 6.98  & 9.74  & 9.58  \\
\hline
Call (SC)\textsuperscript{\textbf{b}}     & 99.85 & 99.40 & 99.70 & 52.63 & 38.44 & 12.87 & 9.71  \\
Call (R50)    & 99.83 & 99.32 & 99.66 & 7.56  & 77.22 & 81.52 & 79.62 \\
\hline
Internet (SC) & 100   & 100   & 100   & 0     & 89.14 & 96.78 & 96.62 \\
Internet (R50)& 99.80 & 99.21 & 99.60 & 0     & 100   & 100   & 100   \\
\hline
Blended (SC)  & 100   & 100   & 100   & 0     & 86.55 & 81.13 & 79.57 \\
Blended (R50) & 100   & 100   & 100   & 0     & 99.35 & 97.22 & 92.79 \\
\hline
\end{tabular}}
{\footnotesize
\textsuperscript{\textbf{a}}ResNet50 SMS values below $10\%$ reflect a fixed-$\tau$ artifact (Sec.~\ref{subsec:calibration}).
\textsuperscript{\textbf{b}}SimpleCNN silent-call Phase~3 decline is interpreted in Sec.~\ref{subsec:gradcam}.}
\end{table}

\subsection{Attack Realism of TabDDPM-Generated Attacks}
\label{subsec:realism_metrics}

We operationalize \emph{realistic} as \textbf{manifold-preserving}: synthesized attacks remain within the statistical envelope of normal CDR traffic, modeling stealthy control-plane abuse rather than high-volume volumetric DDoS. 
We quantify distributional fidelity against the normal CDR aggregate manifold and benchmark against Phase~1 multiplier baselines~\cite{Hussain2021}.

\begin{figure}[t]
\centering
\includegraphics[width=1\linewidth]{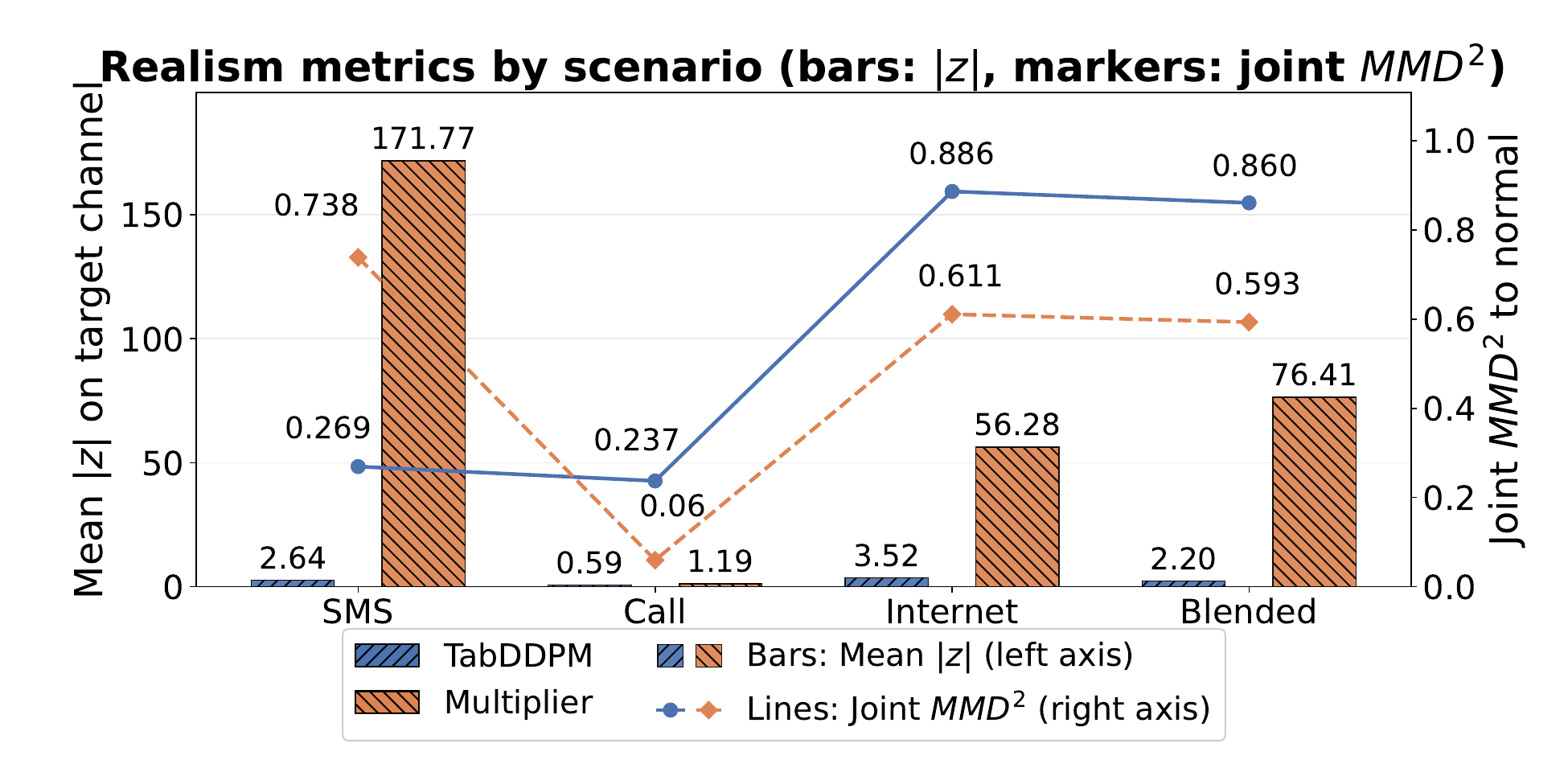}
\caption{Dual-axis realism summary on test attack grids ($168$ per scenario). 
Bars (left axis) show target-channel mean absolute $z$-score, and lines (right axis) show joint (3D) MMD$^2$ to normal, for TabDDPM and multiplier attacks across scenarios. 
Lower values indicate closer alignment with normal CDR aggregates (higher realism/stealth).}
\label{fig:realism_zscore}
\end{figure}

For each test attack grid we extract the mean 3D aggregate over the 40 attacked cells ($168$ grids per scenario) and compare it to a normal reference pool of $5{,}000$ cell-level 3D features from normal-only training grids, 
using target-channel mean absolute $z$-score as the primary realism metric and squared maximum mean discrepancy (MMD$^2$) with a radial basis function (RBF) kernel as a distributional check (full definitions in Supplementary Sec.~S-C).
Separately, $5{,}000$ unconstrained TabDDPM draws yield joint MMD$^2=0.50$ to that normal pool, confirming generative fidelity before scenario constraints. 
After mild scenario constraints, TabDDPM attacks deviate by only $\approx\!0.6$--$3.5\sigma$ on the target channel, whereas multiplier attacks deviate by $\approx\!1$--$172\sigma$ (Fig.~\ref{fig:realism_zscore}; Supplementary Table~SI).
For silent-call, TabDDPM attacks deviate by only $0.59\sigma$ on CALL (target-channel MMD$^2=0.19$), so Phase~2 traffic is statistically near-normal; this stealth regime helps explain why multiplier-trained ResNet50 F1 collapses to $7.6\%$ on TabDDPM Call grids despite near-perfect Phase~1 multiplier scores.
These results confirm TabDDPM as a manifold-preserving stealth-attack synthesizer; limitations are discussed in Sec.~\ref{sec:conclusion}.

\subsection{ADT Recovery and Multiplier Retention} \label{sub:ADT_Robustness_Recovery}

Beyond Table~\ref{tab:phases} (TabDDPM test, protocol~P1), we assess architecture-specific recovery and multiplier-test retention (protocol~P5).

\textbf{SimpleCNN.} On the same $335$-grid TabDDPM protocol (Table~\ref{tab:phases}, $\tau=0.5$), Internet ($0\%\to96.6\%$ at R3) and Blended ($0\%\to79.6\%$) improve markedly.
Silent-call F1 drops from $52.6\%$ (Phase~2) to $9.7\%$ at R3---Grad-CAM analysis (Sec.~\ref{subsec:gradcam}) shows ADT suppresses attack scores on TabDDPM silent-call samples (mean attack score on attacked cells $0.35\to0.01$) while preserving localization; ResNet50 Call F1 still rises to $79.6\%$. SMS F1 remains near $9\%$ throughout.

\begin{figure}[t]
\centering
\includegraphics[width=\columnwidth]{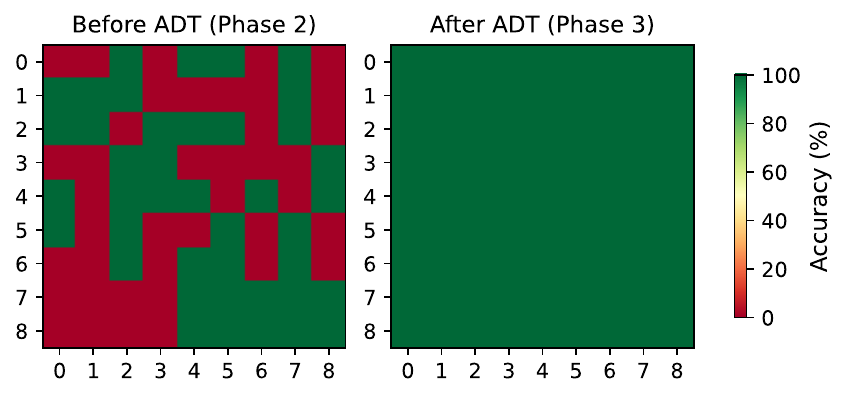}
\caption{Cell-level accuracy heatmaps for the Blended scenario before and after ADT.}
\label{fig:heatmap_blended}
\end{figure}

\textbf{Cell-level localization.} Per-cell accuracy heatmaps for the Blended scenario before and after ADT (Fig.~\ref{fig:heatmap_blended}) show that ADT converts a patchwork of inconsistent per-cell accuracies into a near-uniform $9\times9$ accuracy grid ($\approx 100\%$ per cell), resolving spatial inconsistencies across attacked locations.

\begin{table}[!t]
\centering
\caption{Retention on fixed-multiplier attacks (protocol~P5; ResNet50): F1 (\%) on the locked multiplier test split at $\tau=0.5$, before ADT (Phase~1) and after ADT round~3.}
\label{tab:multiplier_retention}
\small
\begin{tabular}{lcc}
\hline
\textbf{Scenario} & \textbf{Phase 1 (pre-ADT)} & \textbf{ADT round 3} \\
\hline
SMS      & 100.00 & 100.00 \\
Call     &  99.66 & 100.00 \\
Internet &  99.60 &  66.80 \\
Blended  & 100.00 & 100.00 \\
\hline
\end{tabular}
\end{table}

\textbf{Retention on multiplier-based tests.}
Table~\ref{tab:multiplier_retention} shows that ResNet50 retains $100\%$ F1 on SMS, silent-call, and Blended on the \emph{multiplier} test split at $\tau=0.5$ after ADT.
For Internet, F1 decreases from $99.6\%$ to $66.8\%$, while attack-cell recall remains $100\%$ (TP$=6720$, FN$=0$), indicating a calibration shift driven by false positives on normal cells rather than loss of multiplier-attack recognition; increasing $\tau$ to $0.99$ raises Internet F1 to $86.3\%$. 
SimpleCNN retains $100\%$ F1 on all four scenarios after ADT round~3.

\subsection{Comparison with Generative and Adversarial-Training Baselines}
\label{subsec:baseline_comparison}

\begin{figure}[t]
\centering
\includegraphics[width=\columnwidth]{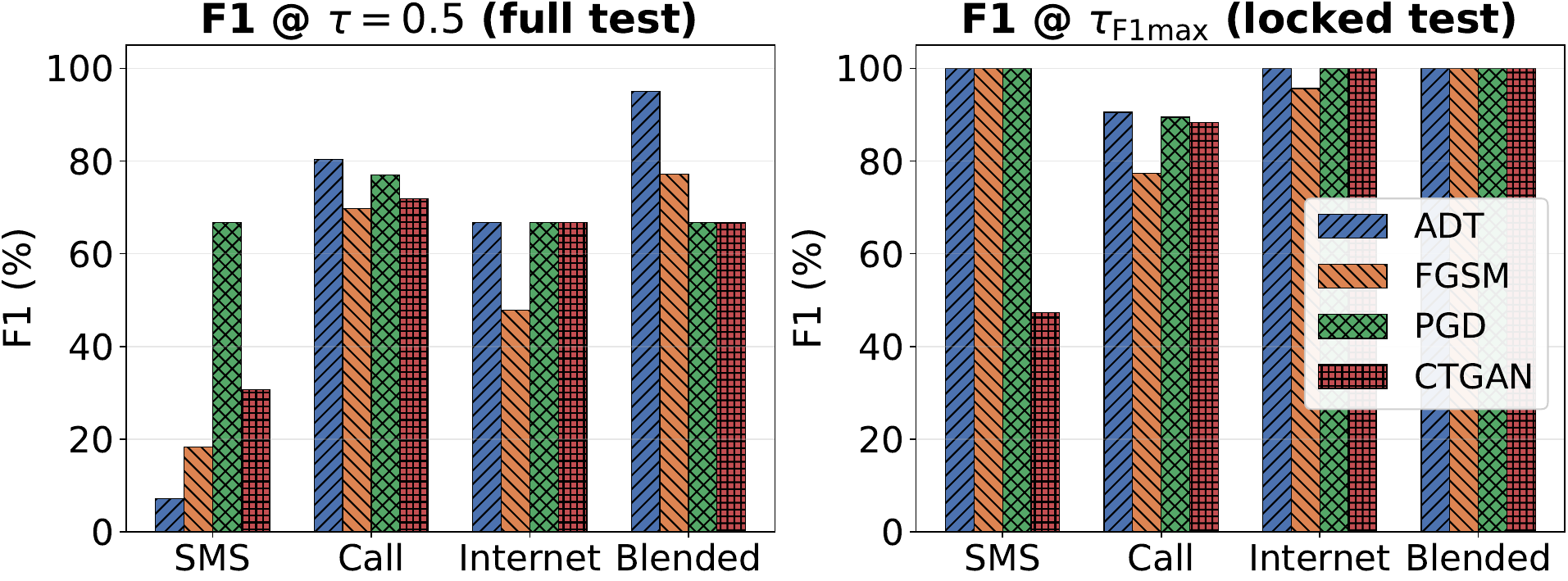}
\caption{Baseline comparison on the locked TabDDPM test set (ResNet50, round~3).
\textbf{Left (P2):} F1 at deployment-default $\tau=0.5$ (FGSM, PGD, and CTGAN each follow the identical 3-round schedule used for ADT). 
\textbf{Right (P3):} F1 at $\tau_{\mathrm{F1max}}$ (validation-selected; Sec.~\ref{subsec:calibration}). 
Compare methods within each panel; each panel applies a single threshold rule. 
Numeric values appear in Supplementary Table~SIII.}
\label{fig:baseline_threshold}
\end{figure}

We compare ADT against a fixed-multiplier-trained detector without iterative retraining~\cite{Hussain2021}, fast gradient sign method (FGSM) adversarial training, projected gradient descent (PGD) adversarial training ($\ell_\infty$, $\varepsilon=0.1\times$, 10 steps), and conditional tabular GAN (CTGAN)~\cite{xu2019modeling} (150 epochs on up to $20{,}000$ normal cell-level CDR features, same mild scenario scaling as Phase~2).
All iterative methods follow the Phase~3 schedule (3 rounds, 100 augmented grids/round, 20 fine-tuning epochs/round) on ResNet50, evaluated on the locked TabDDPM test set (the same $335$-image split as protocol~P1: $167$ normal $+$ $168$ TabDDPM attack grids; indices fixed after Phase~2 generation).
We report $\tau=0.5$ as the deployment-default (attack-type labels are unavailable at deployment time) before presenting validation-calibrated results with per-method $\tau_{\mathrm{F1max}}$ (Sec.~\ref{subsec:calibration}).
Fig.~\ref{fig:baseline_threshold} summarizes round-3 F1 under protocols P2 (left) and P3 (right); compare methods within each panel, as the two panels apply distinct threshold rules (common $\tau=0.5$ versus per-method $\tau_{\mathrm{F1max}}$).

\textbf{Results at $\tau=0.5$ (iterative protocol).}
Fig.~\ref{fig:baseline_threshold} (left panel) and Supplementary Table~SIII report round-3 F1 under protocol~P2 (iterative baseline-comparison: FGSM, PGD, and CTGAN each follow the same 3-round retraining schedule as ADT).
Under this protocol ($\tau=0.5$, locked TabDDPM test), ADT is the only method reaching operational F1 on \textbf{Blended} without per-method tuning ($95.1\%$ vs.\ $\leq77.2\%$ for FGSM/PGD/CTGAN). 
On \textbf{Call}, ADT leads ($80.4\%$; next-best PGD $77.0\%$). On \textbf{Internet}, ADT ties PGD/CTGAN ($66.8\%$).
On \textbf{SMS}, PGD reports higher round-3 F1 at $\tau=0.5$ (P2: $66.8\%$ vs.\ ADT $7.2\%$); on the locked attack test split both achieve AUC$=1.0$---a decision-rule artifact resolved in Sec.~\ref{subsec:calibration} (Fig.~\ref{fig:baseline_threshold}, right panel).

\textbf{Validation-calibrated comparison.}
Fig.~\ref{fig:baseline_threshold} (right panel) reports $\tau_{\mathrm{F1max}}$ selected on the held-out attack validation split and applied once to the locked attack test (P3; distinct from the $335$-grid $\tau=0.5$ scores in Table~\ref{tab:phases}).
ADT and PGD are essentially tied on \textbf{Call} ($90.5\%$ vs.\ $89.5\%$; bootstrap 95\% CIs $[85.6, 94.7]$ and $[83.3, 94.3]$, overlapping). The clearest gap is \textbf{SMS}: ADT/FGSM/PGD reach $100\%$ while CTGAN remains at $47.3\%$. Internet and Blended approach $100\%$ for all methods after calibration.

\begin{table}[!t]
\centering
\caption{Offline generative and augmentation cost (RTX~4090).}
\label{tab:generative_cost}
\small
\setlength{\tabcolsep}{4pt}
\begin{tabular}{lrr}
\toprule
\textbf{Method} & \textbf{Train (s)} & \textbf{Aug.\ 100} \\
\midrule
FGSM / PGD & --- & ${\lesssim}0.01$ \\
CTGAN      & 62.5 & 0.007 \\
TabDDPM    & 1296.0 & --- \\
ADT round  & --- & 21.7 \\
\bottomrule
\end{tabular}
\parbox{\columnwidth}{\small\textit{Note:} FGSM/PGD: no generative model; augmentation is on-the-fly grid perturbation. CTGAN and TabDDPM: one-time offline training. ADT round: inverse-guided TabDDPM generation plus 20-epoch detector fine-tuning per round.}
\end{table}

\textbf{Is diffusion strictly necessary?}
FGSM and PGD are training-time perturbation methods that distort existing grids but cannot synthesize standalone manifold-consistent attack samples for independent stress-testing; PGD is competitive on Call but lacks CDR aggregate realism (Sec.~\ref{subsec:realism_metrics}).
FGSM and PGD require no generative pre-training, whereas CTGAN trains $\approx\!21\times$ faster than TabDDPM ($62.5$\,s vs.\ $1296.0$\,s; Table~\ref{tab:generative_cost}), but showed a large Blended round-wise regression ($99.7\%\to66.8\%$ at $\tau=0.5$) and calibrated SMS F1 of only $47.3\%$ (P3; Supplementary Table~SIII bottom block).
ADT combines manifold-preserving synthesis with deployable Blended robustness at the default threshold---properties not jointly matched by the alternatives.

\begin{table}[t]
\centering
\caption{Inverse-guidance sensitivity summary (ResNet50, round-3 F1\,\%).
The operational default row reports P1 (cumulative Phase-3 ADT on the full $335$-grid TabDDPM test, Table~\ref{tab:phases});
sweep rows report a fresh-retrain, locked-test one-at-a-time sweep.
The full grid is in Supplementary Table~SII.}
\label{tab:guidance_sensitivity}
\small
\setlength{\tabcolsep}{4pt}
\begin{tabular}{lccccc}
\hline
\textbf{Protocol} & $\lambda$ & $p_{\mathrm{t}}$ & \textbf{Blended} & \textbf{SMS} & \textbf{Internet} \\
\hline
Operational default & 1.0 & 0.3 & 92.8 & 9.6 & 100.0 \\
\hline
\multirow{3}{*}{Fresh-retrain sweep} & 1.0 & 0.3 & 66.8 & 100.0 & 66.8 \\
 & 2.0 & 0.3 & 66.8 & 100.0 & 100.0 \\
 & 1.0 & 0.1 & 100.0 & 100.0 & 65.9 \\
\hline
\end{tabular}
\end{table}

\begin{table}[t]
\centering
\caption{Online single-grid inference latency for Phase~3 (ADT) detectors (batch~$=1$, scenario-averaged; RTX~4090, 500 iterations after 50 warmup).
The per-scenario and batch-$=32$ breakdown is in Supplementary Table~SVIII.}
\label{tab:inference_latency}
\small
\begin{tabular}{lrrr}
\toprule
\textbf{Arch.} & \textbf{Mean (ms)} & \textbf{p95 (ms)} & \textbf{Grids/s} \\
\midrule
SimpleCNN & 35.7 & 49.5 & 28.1 \\
ResNet50  & 42.7 & 60.4 & 23.4 \\
\bottomrule
\end{tabular}
\end{table}

\subsection{Inverse-Guidance Sensitivity and Online Latency}
\label{subsec:guidance_sensitivity} \label{subsec:inference_latency}
ADT depends on guidance scale $\lambda$ (steering strength) and $p_{\mathrm{target}}$ (adversarial difficulty). 
We perform a one-at-a-time sweep on ResNet50: $\lambda\in\{0.25,0.5,1.0,2.0\}$ at $p_{\mathrm{target}}{=}0.3$, and $p_{\mathrm{target}}\in\{0.1,0.3,0.5,0.7\}$ at $\lambda{=}1.0$. 
Target-channel MMD$^2$ stays flat ($\approx0.51$--$0.57$), so neither hyperparameter materially affects on-manifold realism. 
On the fresh-retrain one-at-a-time sweep (P4; locked TabDDPM attack test, round-3 F1 at $\tau=0.5$) in Table~\ref{tab:guidance_sensitivity}, robustness is \emph{sensitive and non-monotonic}: 
Internet F1 rises from $66.8\%$ at $\lambda{=}1.0$ to $100\%$ at $\lambda{=}2.0$; Blended F1 rises from $66.8\%$ at $p_{\mathrm{target}}{=}0.3$ to $100\%$ at $p_{\mathrm{target}}{=}0.1$.
The table also reports the operational Phase-3 default; the full sweep is in Supplementary Sec.~S-D.
$(1.0, 0.3)$ is the fixed deployment compromise, not a global optimum.

Runtime CPS monitoring requires only the online forward-pass latencies in Table~\ref{tab:inference_latency}.
Diff-DDoS targets \emph{periodic} CDR grid analytics, not per-packet or Ultra-Reliable and Low-Latency Communications (uRLLC)-level ($\sim$1\,ms) inference: each $9\times9\times3$ grid aggregates a \textbf{10-minute} window (Sec.~\ref{sec:experimental_setup}), so the operative CPS latency budget is the inter-grid interval ($600$\,s).
SimpleCNN attains $35.7$\,ms mean ($49.5$\,ms p95) and ResNet50 $42.7$\,ms mean ($60.4$\,ms p95) on an RTX~4090 reference GPU; worst batch-$1$ p95 is $\approx\!10^{4}\times$ below the aggregation window (Supplementary Table~SVIII).
Offline pipeline cost is in Supplementary Sec.~S-H (Table~SVI).

\vspace{-4pt}
\subsection{Operating Points: ROC-AUC and Threshold Calibration} \label{subsec:roc_auc} \label{subsec:calibration}

\begin{figure}[t]
\centering
\includegraphics[width=1\columnwidth]{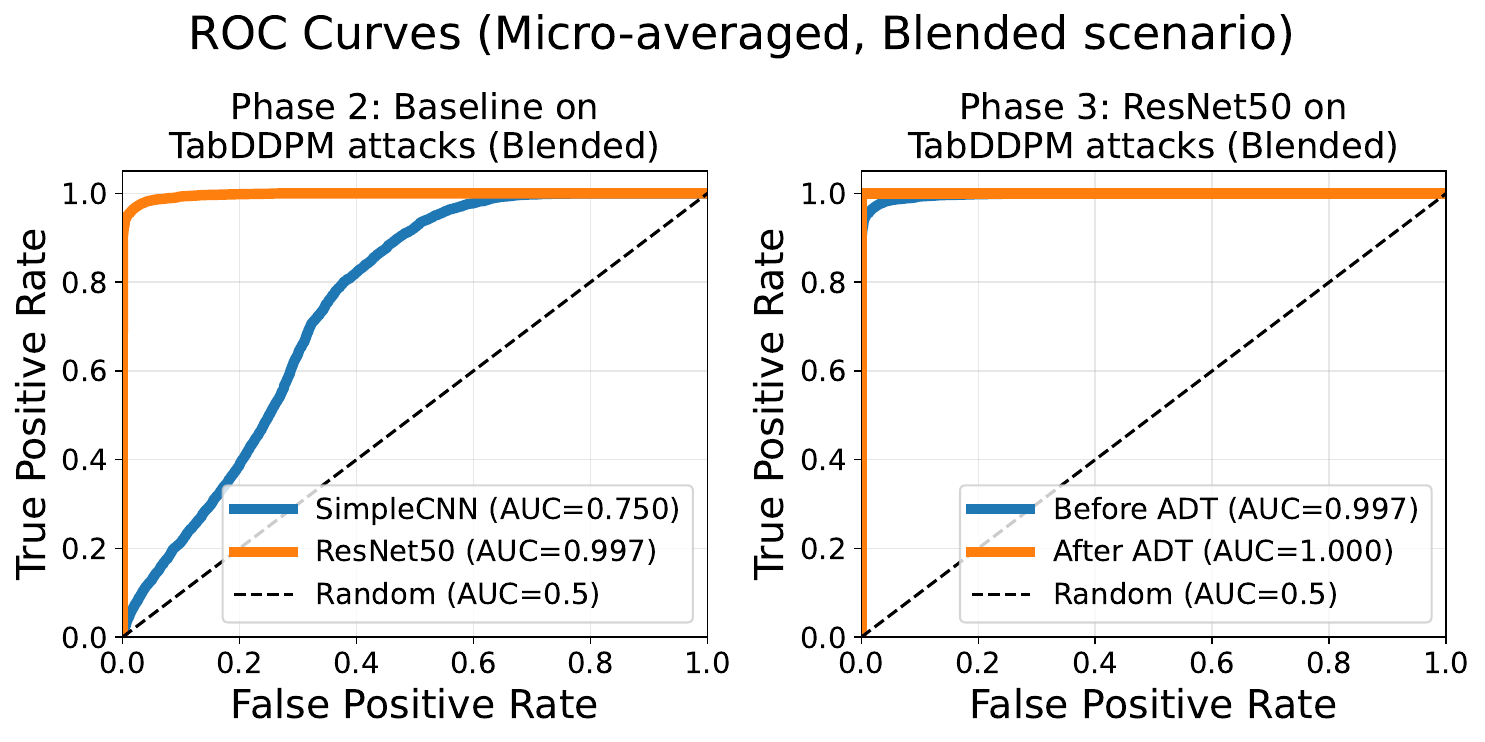}
\caption{ROC curves for the Blended scenario on TabDDPM-generated attacks. 
Left: Phase 2 comparison of SimpleCNN and ResNet50. Right: ResNet50 before and after ADT in Phase 3.}
\label{fig:roc_blended}
\end{figure}

\begin{figure*}[!t]
\centering
\includegraphics[width=2\columnwidth]{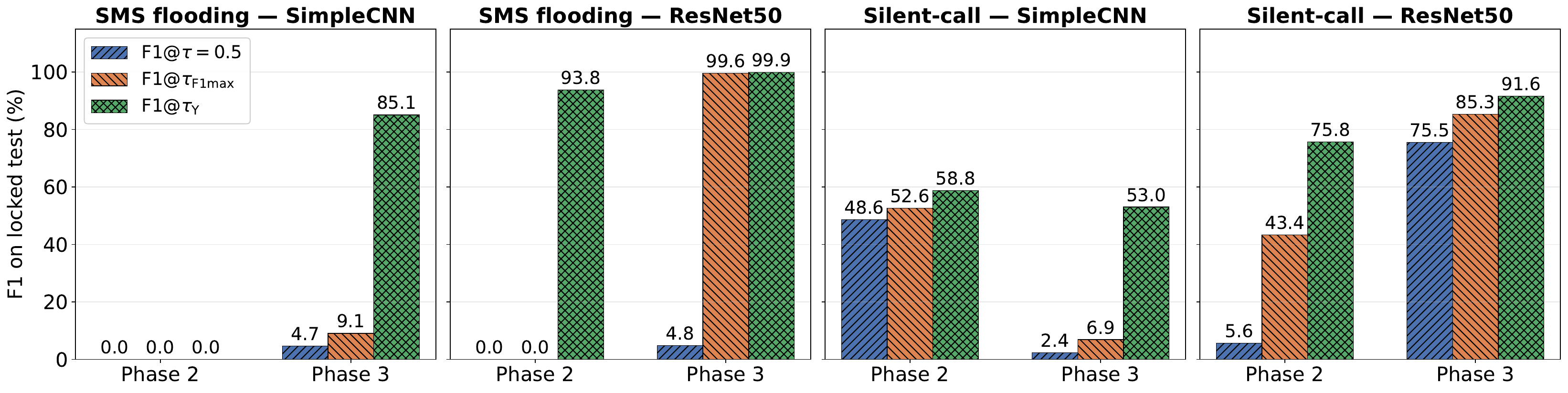}
\caption{Locked-test F1 at the default threshold versus validation-calibrated thresholds ($\tau_{\mathrm{F1max}}$, $\tau_{\mathrm{Y}}$) for the SMS and silent-call scenarios.}
\label{fig:calibration}
\vspace{-12pt}
\end{figure*}

Micro-averaged ROC~AUC (full table in Supplementary Table~SVII; representative Blended ROC curves in Fig.~\ref{fig:roc_blended}, with additional ROC/PR curves in the Supplement) confirms Phase~1 perfect separability (AUC$=1.0$). Phase~2 reveals vulnerability: SimpleCNN drops to AUC$=0.43$ (SMS) and $0.75$ (Blended), while ResNet50 holds $0.90$--$0.99$. After ADT, SimpleCNN recovers to $0.85$ (SMS) and $0.96$ (Blended); ResNet50 reaches $0.95$--$1.0$.
The AUC/F1 gap is a decision-rule artifact: for SMS the detector correctly ranks attacks (ResNet50 AUC up to $1.0$ after ADT) 
yet $\tau=0.5$ yields near-zero SMS F1 (Table~\ref{tab:phases}; locked attack split in Fig.~\ref{fig:calibration}) because attack scores concentrate below $0.5$; validation-based threshold calibration (below) resolves this.

We implement validation-based threshold calibration and cost-sensitive retraining for the two threshold-sensitive sparse-attack scenarios (SMS flooding and silent-call).
For each, $\tau_{\mathrm{F1max}}$ (validation F1-maximizing) and $\tau_{\mathrm{Y}}$ (Youden's $J=\mathrm{TPR}-\mathrm{FPR}$)~\cite{youden1950} are selected on a held-out validation split of the TabDDPM \emph{attack-only} test grids ($84$ of $168$ attack images, $50\%$ split, seed~$1$) and applied once to the remaining $84$ locked attack test grids (separate from the primary $335$-image micro-averaged evaluation in Table~\ref{tab:phases}), so no test-set tuning occurs.
Fig.~\ref{fig:calibration} compares locked-test F1 at $\tau=0.5$, $\tau_{\mathrm{F1max}}$, and $\tau_{\mathrm{Y}}$ for Phase~2 (TabDDPM, no ADT) versus Phase~3 (ADT) on SMS flooding and silent-call.
For ResNet50 on SMS flooding, calibration alone resolves the apparent ADT failure: on the locked attack split (Fig.~\ref{fig:calibration}), F1 rises from $4.82\%$ at $\tau=0.5$ to $99.55\%$ at $\tau_{\mathrm{F1max}}=0.05$ (AUC$=1.0$, FPR$=0\%$; Table~SIV), confirming that ADT learned the attack and that the sub-$10\%$ SMS F1 at $\tau=0.5$ in Table~\ref{tab:phases} was a fixed-threshold artifact rather than a learning failure.

We also evaluate class-weighted BCE with positive-class weight ($w_{+}\!\approx\!8.5$) and focal loss ($\gamma=2$, $\alpha=0.75$)~\cite{lin2020focal} over five seeds.
For SMS flooding, neither calibration nor cost-sensitive retraining raises SimpleCNN F1 above $12.3\pm8.6\%$ (large seed variance), indicating a capacity limit on the lightweight backbone; we therefore adopt ResNet50 as the deployable detector for sparse attacks.
For silent-call, weighted BCE raises mean locked-test AUC from $0.65$ to $0.85\pm0.09$, but F1 at $\tau_{\mathrm{F1max}}$ remains only $48.5\pm25.3\%$ with large seed variance.
The full per-architecture grid, Youden thresholds with FPR, and cost-sensitive comparisons appear in Supplementary Sec.~S-F (Table~SIV); see also Fig.~\ref{fig:calibration}.

\subsection{Grad-CAM Interpretability of the SimpleCNN Silent-Call Anomaly}
\label{subsec:gradcam}

\begin{figure}[!t]
\centering
\includegraphics[width=\columnwidth]{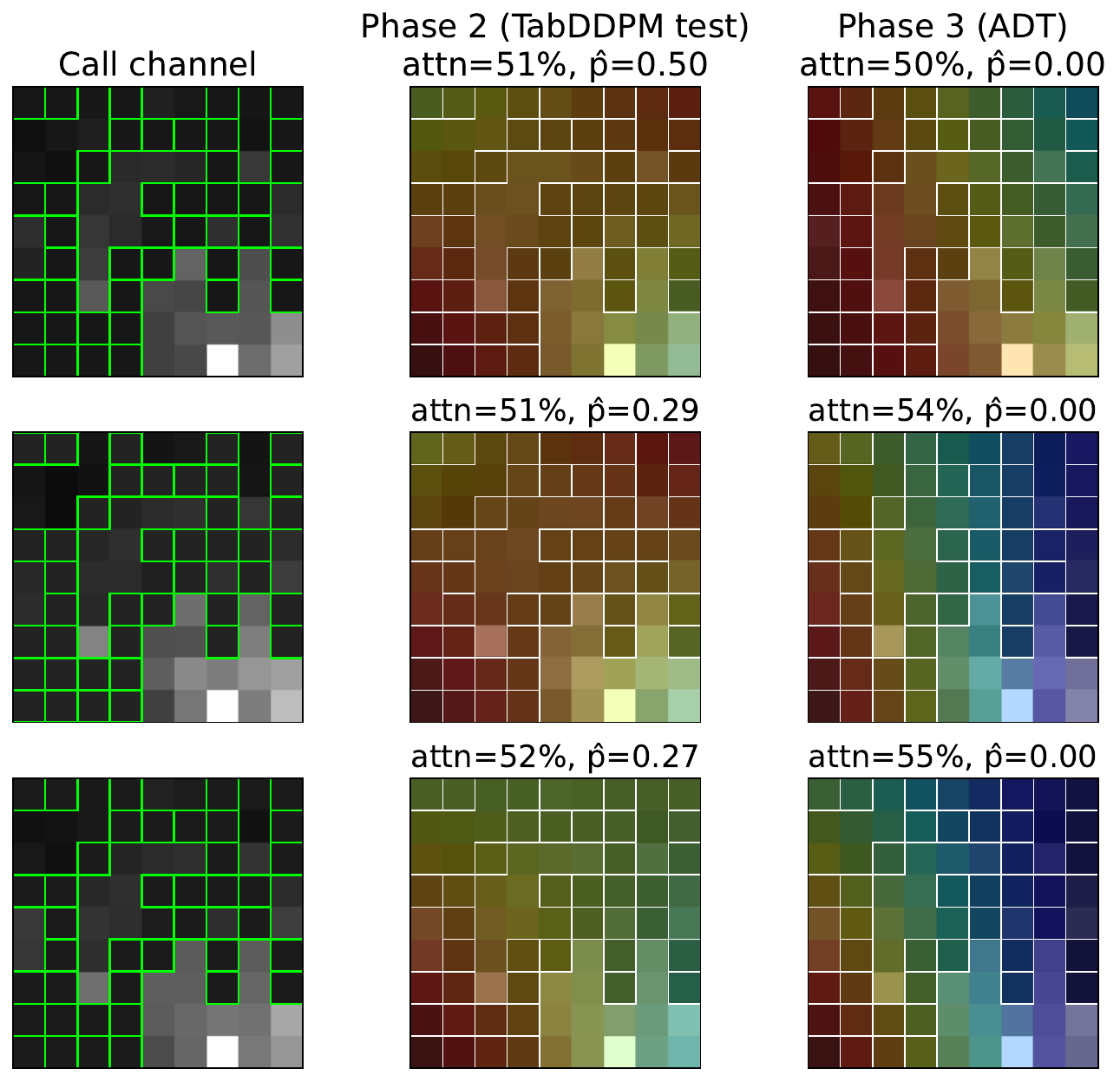}
\caption{Grad-CAM overlays for TabDDPM silent-call attacks (SimpleCNN): Call-channel input (left) and saliency maps for Phase~2 (trained on multiplier attacks, tested on TabDDPM silent-call grids) vs.\ Phase~3 ADT (same backbone after retraining). 
Boxed cells mark ground-truth attacked locations (green on the input panel, white on the overlays). 
Subplot titles report the mean predicted attack score on attacked cells ($\hat{p}$) and the fraction of Grad-CAM saliency mass on those cells (\textsf{attn}).}
\label{fig:gradcam_call}
\end{figure}

SimpleCNN silent-call F1 falls from $52.6\%$ (Phase~2) to $9.7\%$ (Phase~3, Table~\ref{tab:phases}), which appears to contradict the ADT robustness claim.
We apply Gradient-weighted Class Activation Mapping (Grad-CAM)~\cite{selvaraju2020gradcam} to SimpleCNN's last convolutional layer on three grid types---normal (real) CDR snapshots, TabDDPM silent-call attacks, and fixed-multiplier baseline attacks---comparing the Phase~2 detector with the Phase~3 ADT model over $N{=}40$ test grids per condition. 
Warmer overlay colors in Fig.~\ref{fig:gradcam_call} indicate stronger Grad-CAM activation (where the model was most sensitive on that grid image); $\hat{p}$ is the mean predicted attack score on attacked cells, analogous to $p_{\mathrm{attack}}$ in Sec.~\ref{sec:phase3}.
The figure shows three illustrative TabDDPM silent-call grids in which Phase~2 assigns moderate $\hat{p}$ on attacked cells while Phase~3 drives $\hat{p}$ near zero, even though the fraction of Grad-CAM mass on attacked cells (\textsf{attn}) stays $\approx 50\%$ in both phases.

\begin{figure}[!t]
\centering
\includegraphics[width=1\columnwidth]{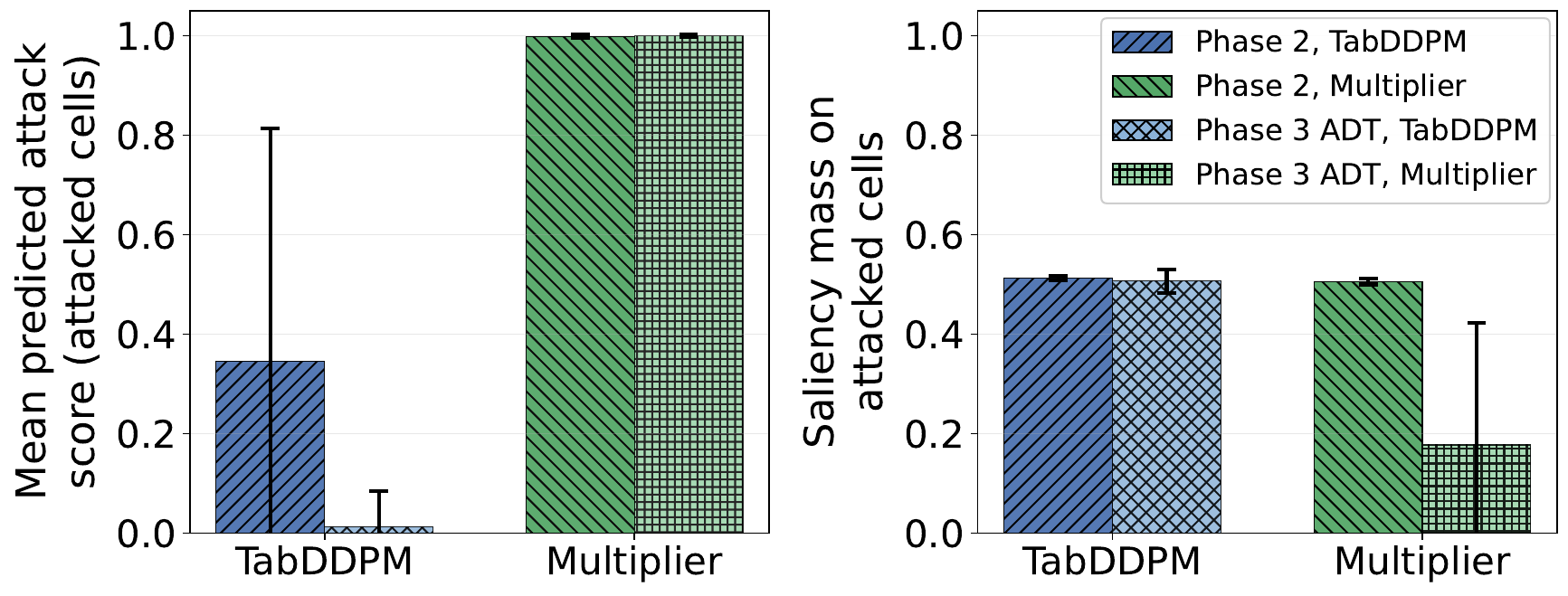}
\caption{Quantitative Grad-CAM summary for SimpleCNN before vs.\ after ADT on attacked cells.
\textit{Left:} mean predicted attack score on attacked cells for TabDDPM and multiplier attack sets.
\textit{Right:} Grad-CAM saliency mass on attacked cells for the same sets.
Bars compare Phase~2 and Phase~3 (ADT); error bars denote $\pm 1$ sample standard deviation over $N=40$ test grids per condition.}
\label{fig:gradcam_compare}
\end{figure}

Fig.~\ref{fig:gradcam_compare} summarizes the score--saliency asymmetry quantitatively: across $N{=}40$ TabDDPM silent-call test grids, mean attacked-cell scores drop from $0.35\pm0.47$ (Phase~2) to $0.01\pm0.07$ (Phase~3), matching the F1 collapse, while saliency mass on attacked cells stays comparable ($0.51\pm0.01$ vs.\ $0.51\pm0.02$).
On multiplier-based attacks, mean scores remain near $1.0$ in both phases, while Phase~3 saliency becomes diffuse ($0.18\pm0.24$ vs.\ $0.51\pm0.01$; Supplementary Table~SV).
ADT on SimpleCNN therefore overfits to stealth guidance samples and \emph{suppresses} detection on manifold-preserving TabDDPM traffic rather than losing localization; ResNet50 still reaches $79.6\%$ F1 under the same protocol. 
Full numerical Grad-CAM results are in Supplementary Sec.~S-G (Table~SV).

\section{Limitations and Conclusion}\label{sec:conclusion}
We found that multiplier-trained CDR detectors fail on manifold-preserving attacks: Phase~2 F1 collapsed to near zero in three of four scenarios (Table~\ref{tab:phases}), while ADT with ResNet50 recovered deployable F1 for silent-call ($79.6\%$), Internet ($100\%$), and Blended ($92.8\%$), retaining near-perfect multiplier-test performance except an Internet calibration shift (Table~\ref{tab:multiplier_retention}).
This extends Hussain \emph{et al.}~\cite{Hussain2021} from multiplier baselines to manifold-preserving stress tests and applies inverse-guided TabDDPM for iterative cellular IDS hardening.
Calibration resolves the SMS threshold artifact (Sec.~\ref{subsec:calibration}), Grad-CAM shows SimpleCNN can suppress stealth silent-call scores without losing localization (Sec.~\ref{subsec:gradcam}), and ADT matched or exceeded gradient- and GAN-based baselines (Sec.~\ref{subsec:baseline_comparison}).

Three limitations bound external validity.
No labeled per-cell cellular DDoS CDR ground truth exists (Sec.~\ref{subsec:eval_scope}), so evaluation uses synthetic rather than operator-confirmed attack traces.
Attack realism uses coarse 3D aggregate statistics---mean $|z|$ and MMD$^2$ on SMS/CALL/INTERNET sums (Sec.~\ref{subsec:realism_metrics})---a proxy that omits full grid spatiotemporal structure and does not establish field adversary fidelity.
Finally, the Milano corpus reflects 3G/4G-era metropolitan traffic~\cite{barlacchi2015multi}, and ten-minute grid monitoring targets periodic CPS analytics rather than sub-millisecond uRLLC control (Sec.~\ref{subsec:inference_latency}).

Operator CDR validation and packet-level 5G corpora where mapping permits~\cite{siriwardhana2025data,ncsrd2024_5gddos} should confirm ADT transfer beyond Milano; future work includes adaptive adversaries and regional scale-up~\cite{hussain2026closedloop}.
For 5G-connected CPS and IoT services, stealthy signaling abuse threatens network and physical-process availability; pre-deployment stress testing is a practical safeguard.
Manifold-preserving TabDDPM attacks expose multiplier-trained brittleness, and ADT recovers robustness with ResNet50 and calibration, offering a workflow to harden cell-level DDoS monitors when labeled attack CDRs remain scarce.

\bibliographystyle{IEEEtran}
\bibliography{references}

\begin{IEEEbiography}[{\includegraphics[width=1in,height=1.25in,clip,keepaspectratio]{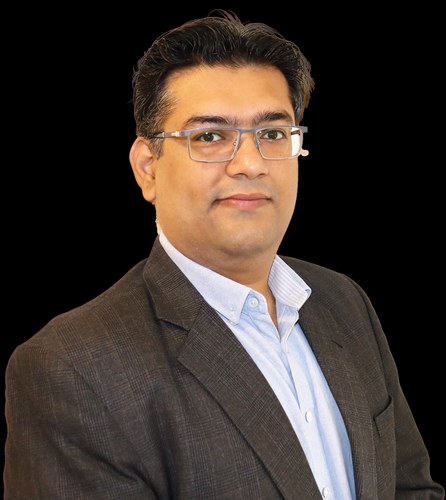}}]{Bilal Hussain}
(Senior Member, IEEE) received the B.E. degree in electrical engineering from Bahria University, Pakistan, in 2010, the M.Sc. degree in information and communications engineering from the University of Leicester, Leicester, U.K., in 2011, and dual Ph.D. degrees in information and communications engineering from The Hong Kong Polytechnic University (PolyU), Hong Kong SAR, China, in 2021, and Xi'an Jiaotong University, Xi'an, China, in 2022. He is currently a Lecturer with the Division of Science, Engineering, and Health Studies, School of Professional Education and Executive Development, PolyU. Previously, he was a Postdoctoral Fellow with the Centre for Advances in Reliability and Safety, Hong Kong. His research focuses on intelligent 6G networks, Agentic AI, real-time edge intelligence, and robust cyber-physical systems security.
\end{IEEEbiography}

\begin{IEEEbiography}[{\includegraphics[width=1in,height=1.25in,clip,keepaspectratio]{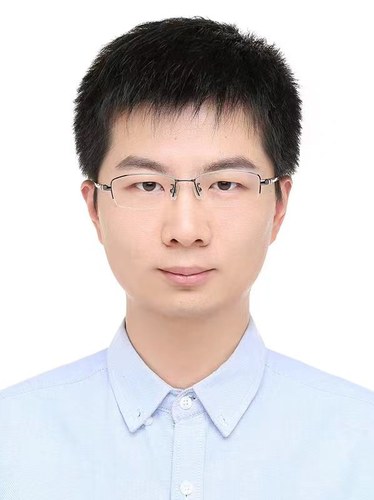}}]{Xiao Tang}
(Member, IEEE) received the B.S. degree in information engineering (Elite Class Named After Tsien Hsue-shen) and the Ph.D. degree (Hons.) in information and communication engineering from Xi'an Jiaotong University, Xi'an, China, in 2011 and 2018, respectively. He is currently an Associate Professor with the School of Information and Communication Engineering, Xi'an Jiaotong University, Xi'an, China.
\end{IEEEbiography}

\begin{IEEEbiography}[{\includegraphics[width=1in,height=1.25in,clip,keepaspectratio]{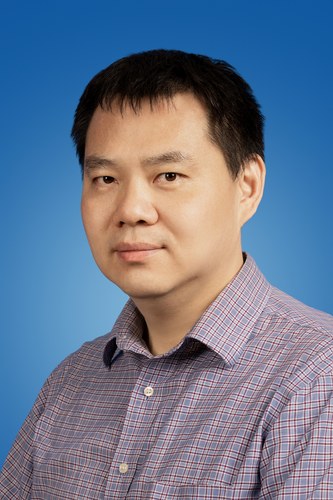}}]{Qinghe Du}
(Member, IEEE) 
received the B.S. and M.S. degrees in electrical engineering from Xi'an Jiaotong University, China, in 2001 and 2004, respectively, and the Ph.D. degree in computer engineering from Texas A\&M University, College Station, TX, USA, in 2010. He is currently a Professor with the School of Information and Communications Engineering, Xi'an Jiaotong University. He has published over 150 technical papers in refereed technical journals and international academic conferences. His research interests widely cover related areas of mobile wireless communications and networking. He received several Best Paper Awards from the international academic conferences and technical journals. He served and/or serves as an Associate Editor for the \textsc{IEEE Transactions on Communications} and \textsc{IEEE Communications Letters}, an Area Editor for \textit{KSII Transactions on Internet and Information Systems}, and an Editor for \textit{Electronics}.
\end{IEEEbiography}

\begin{IEEEbiography}[{\includegraphics[width=1in,height=1.25in,clip,keepaspectratio]{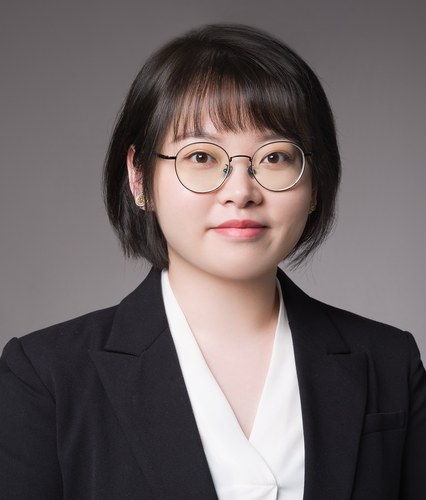}}]{Tan Li}
(Member, IEEE) received the Ph.D. degree in computer science from City University of Hong Kong, Hong Kong SAR, China, in 2023, the M.Eng. degree in control engineering from the University of Science and Technology of China, Hefei, China, in 2019, and the B.Eng. degree in automation from Central South University, Changsha, China, in 2016. She is currently an Assistant Professor with the Department of Computer Science, The Hang Seng University of Hong Kong, Hong Kong SAR, China. Her research interests include federated learning and machine learning for wireless communications.
\end{IEEEbiography}

\begin{IEEEbiography}[{\includegraphics[width=1in,height=1.25in,clip,keepaspectratio]{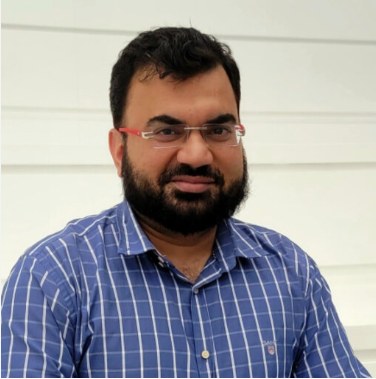}}]{Muhammad Azhar} 
(Member, IEEE) received the B.S. degree in computer science from the National University of Computer and Emerging Sciences, Islamabad, Pakistan, the master's degree in computer science from Sejong University, Seoul, South Korea, and the Ph.D. degree in machine learning and data mining from Shenzhen University, Shenzhen, China. He is currently an Associate Professor and the Associate Head of the Department of Applied Data Science with Hong Kong Shue Yan University, Hong Kong SAR, China. He also serves as the Director of the university's Big Data Laboratory. 
His research interests include data mining, natural language processing, and advanced deep-learning techniques.
\end{IEEEbiography}

\begin{IEEEbiography}[{\includegraphics[width=1in,height=1.25in,clip,keepaspectratio]{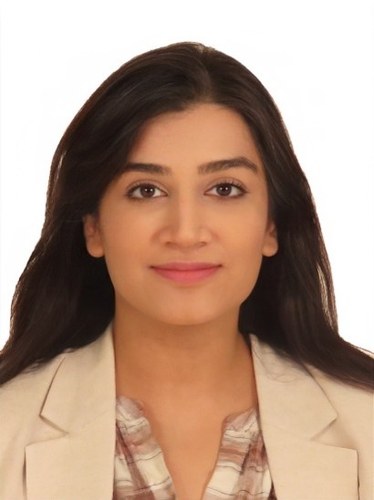}}]{Danista Khan}
(Member, IEEE) received the Ph.D. degree in electrical and electronic engineering from The Hong Kong Polytechnic University (PolyU), Hong Kong SAR, China, specializing in applied deep-learning methods for wireless sensing. She received the master's degree in electrical engineering from the University of Engineering and Technology, Lahore, Pakistan, and the bachelor's degree in electrical engineering from The University of Lahore, Lahore, Pakistan. 
She is currently a Lecturer with the Division of Business and Hospitality Management, School of Professional Education and Executive Development, PolyU. Her research interests include wireless communications and networking, wireless sensing, and the Internet of Things.
\end{IEEEbiography}

\end{document}